\documentclass[showpacs,preprintnumbers,twocolumn,amsmath,amssymb,floatfix]{revtex4}

\usepackage{graphics}
\usepackage{epsfig}

\begin{document}

\title{Quark Potential in a Quark-Meson Plasma}
\author{\normalsize{Chengfu Mu and Pengfei Zhuang}}
\affiliation{Physics Department, Tsinghua University, Beijing
100084, China}

\begin{abstract}
We investigate quark potential by considering meson exchanges in
the two flavor Nambu--Jona-Lasinio model at finite temperature and
density. There are two kinds of oscillations in the chiral
restoration phase, one is the Friedel oscillation due to the sharp
quark Fermi surface at high density, and the other is the Yukawa
oscillation driven by the complex meson poles at high temperature.
The quark-meson plasma is strongly coupled in the temperature
region $1\le T/T_c \lesssim 3$ with $T_c$ being the critical
temperature of chiral phase transition. The maximum coupling in
this region is located at the critical point.
\end{abstract}
 \maketitle

\section
{introduction}
\label{s1}

It is widely accepted that there are two QCD phase transitions at
finite temperature and density, one is the deconfinement and the
other is the chiral symmetry restoration. The only possible way to
realize the phase transitions in laboratories is through high
energy nuclear collisions. From the experimentally observed
anisotropic collective flow at RHIC which is characterized by the
perfect fluid, the surviving resonant states of quarks and gluons
above the critical temperature $T_c$ of the phase transitions
calculated in lattice QCD and effective models, and the
significant difference between the lattice calculated
thermodynamic functions and the corresponding Stefan-Boltzmann
limits at extremely high temperature, the created new state of
matter at RHIC is a strongly coupled quark-gluon
plasma~\cite{sqgp}. With the free energy of a pair of heavy quarks
calculated in lattice QCD~\cite{lattice}, the heavy quark
potential can be extracted. The depth of the potential well
decreases with increasing temperature and the dissociation
temperature of the bound state $J/\Psi$ is about $2.7\
T_c$~\cite{shuryak1}.

From Yukawa's idea that nucleon-nucleon scattering is through
boson exchanges, the static nucleon potential is a Fourier
transformation of the boson propagator at momentum
$K_\mu=(k_0=0,{\bf k})$,
\begin{equation}
\label{potential}
V(r) = \int{d^3{\bf k}\over (2\pi)^3}e^{i{\bf
k}\cdot{\bf r}}U(0,k^2).
\end{equation}
The Yukawa potential is widely discussed in nuclear matter and
recently extended to quark matter, especially, its two kinds of
oscillations, the Friedel oscillation~\cite{kapusta, alonso1,
durso, alonso2, sivak1, epelbaum} induced by the sharp Fermi
surface at high density and the Yukawa
oscillation~\cite{sivak1,shuryak2} resulted from the complex poles
of the exchanged bosons, are frequently emphasized.

To understand the quark behavior in strongly coupled quark matter
with quarks, gluons and their bound states as constituents, we
investigate in this paper the light quark potential through meson
exchanges in a quark-meson plasma controlled by chiral dynamics.
By calculating the quark potential around the chiral phase
transition, we hope to see the effect of chiral symmetry
restoration on the quark properties, to extract the information on
the temperature region of strong coupling, and to know where the
maximum coupling is.

In our calculation we construct mesons in quark matter in the
frame of Nambu--Jona-Lasinio~\cite{njl} (NJL) model at quark
level~\cite{qnjl}. The NJL model is inspired from the electron
superconductivity. It describes well the chiral properties in
vacuum and in hot and dense medium. Above the critical
temperature, there exist hadronic excitations with small
widths~\cite{hatsuda, zhuang}. The spatial dependence of static
meson correlation functions in the model was investigated at
finite temperature~\cite{florkowski1} and
density~\cite{florkowski2}. In recent years, the model is
successfully used to study color superconductivity at moderate
baryon density~\cite{csc} and pion superfluidity at finite isospin
density~\cite{pion}. To understand the results of lattice QCD
thermodynamics in terms of quasiparticle degrees of freedom, the
model is recently extended to include Polyakov loop dynamics
(PNJL)~\cite{pnjl1, pnjl2}.

To define the NJL model completely as an effective theory, a
regularization scheme must be specified to deal with the important
integrals that occur. A regularization scheme specifies a length
scale for the theory, which is normally expressed as a cutoff
$\Lambda$ on the quark momentum and meson momentum. From the
uncertainty principle, the length scale in the NJL model is $R\sim
1/\Lambda$. Taking the standard cutoff value $\Lambda \sim 600$
MeV~\cite{qnjl}, we have $R \sim 1/3$ fm. This means that, the
quark potential calculated in the NJL model might be reasonable in
the region of $r>1/3$ fm. In the short range the model is probably
not applicable. While we will take a covariant Pauli-Villars
regularization scheme which allows us to do the momentum
integrations in the whole region without an explicit cutoff, we
still focus only on the long range behavior of the quark potential
with $r>1/3$ fm.

The paper is organized as follows. We review the analytic
properties of meson polarization functions in the Pauli-Villars
regularization scheme and determine the meson poles in the complex
momentum plane in Section \ref{s2}. In Section \ref{s3} we show
the quark potential at finite temperature and density and analyze
the effect of chiral symmetry restoration. The temperature region
of strongly coupled quark-meson plasma and the position
corresponding to the maximum coupling will be discussed in this
section. We conclude in Section \ref{s4}.

\section {quark potential}
\label{s2}

The flavor SU(2) NJL model is defined through the lagrangian
density
\begin{equation}
\label{njl}
{\cal L}=\bar{\psi}
\left(i\gamma^\mu\partial_\mu-m_0+\mu\gamma_0\right)\psi
+G\left[\left(\bar{\psi}\psi\right)^2+\left(\bar{\psi}i\gamma_5\tau\psi\right)^2\right]
\end{equation}
with scalar and pseudoscalar interactions corresponding to
$\sigma$ and $\pi$ excitations respectively, where $m_0$ is the
current quark mass, $G$ is the four-quark coupling constant with
dimension (GeV)$^{-2}$, $\tau_i\ (i=1,2,3)$ are the Pauli matrices
in flavor space, and $\mu$ is the quark chemical potential which
is one third of the baryon chemical potential $\mu_B$.

In mean field approximation the inverse quark propagator can be
written as
\begin{equation}
\label{quark}
S^{-1}(P)= \gamma^\mu P_\mu-M+\gamma^0\mu,
\end{equation}
where the dynamically generated quark mass $M$ in the medium is
related to the order parameter $\langle\bar\psi\psi\rangle$ of the
chiral phase transition through
$M=m_0-2G\langle\bar{\psi}\psi\rangle$ and is determined by the
self-consistent gap equation,
\begin{equation}
\label{gap1}
M = m_0+2iG\int{d^4P\over
(2\pi)^4}{\text{Tr}}\left[S(P)\right],
\end{equation}
where and in the following the trace is taken in color, flavor and
Dirac spaces. After the summation over Matsubara fermion
frequencies in the imaginary formalism of finite temperature field
theory, the gap equation can be reexpressed as
\begin{equation}
\label{gap2}
M(1-2GI_1)=m_0.
\end{equation}

Since the NJL model is non-renormalizable, one should take a
regularization scheme to avoid the divergence in the momentum
integration (\ref{gap1}). By taking into account the covariant
Pauli-Villars regularization approach, we can separate the
function $I_1$ in (\ref{gap2}) into a vacuum and a matter
part~\cite{florkowski1},
\begin{eqnarray}
\label{i1}
&& I_1 = I_1^{vac}+I_1^{mat},\nonumber\\
&& I_1^{vac} = {3\over 2\pi^2}\sum_{j=0}^N A_j\Lambda_j^2\ln
\Lambda_j^2,\nonumber\\
&& I_1^{mat} = -{6\over \pi^2}\sum_{j=0}^N\sum_{a=\pm}
A_j\int_0^\infty dp{p^2\over E_j(p)}f\left(E_j^a(p)\right)\ \ \
\end{eqnarray}
with the definition of $E_j^\pm(p)=E_j(p)\pm\mu$ and $E_j(p)
=\sqrt{p^2+\Lambda_j^2}$, the Fermi-Dirac distribution function
$f(x)=1/\left(e^{x/T}+1\right)$ at finite temperature $T$, and the
number of subtractions $N$. In the following numerical
calculations we take the same parameter values as used in Ref.
~\cite{florkowski1}, $m_0=8.56$ MeV, $G=0.75/2$ fm$^{2}$, $N=3$,
$\Lambda_0=M$, $\Lambda_1=680$ MeV, $\Lambda_2=2.1\Lambda_1,
\Lambda_3=2.1\Lambda_2$ and $A_0=1$, and $A_1, A_2, A_3$ are
determined by the constraints $\sum_{j=0}^N A_j=0$, $\sum_{j=0}^N
A_j\Lambda_j^2=0$ and $\sum_{j=0}^N A_j\Lambda_j^4=0$. With these
values one can fit the pion mass $M_\pi=138$ MeV and pion decay
constant $f_\pi=94$ MeV and fix the quark mass $M=376$ MeV and
sigma mass $M_\sigma=760$ MeV in the vacuum.

In the NJL model, mesons are considered as quantum fluctuations
above the mean field and can be constructed in random phase
approximation (RPA). After the bubble summation in RPA, the meson
propagator is expressed as~\cite{qnjl,zhuang}
\begin{equation}
\label{meson1}
U_m(k_0^2,k^2)={-2G\over 1-2G\Pi_m(k_0^2,k^2)}
\end{equation}
with the meson polarization functions
\begin{equation}
\label{meson2}
\Pi_m(k_0^2,k^2) = i\int{d^4P\over (2\pi)^4} {\text
{Tr}} \left[\Gamma_m^* S(P+{K\over 2})\Gamma_{n} S(P-{K\over
2})\right],
\end{equation}
and the meson vertexes
\begin{equation}
\label{vertex}
\Gamma_m = \left\{\begin{array}{ll}
1 & m=\sigma\\
i\tau_+\gamma_5 & m=\pi_+ \\
i\tau_-\gamma_5 & m=\pi_- \\
i\tau_3\gamma_5 & m=\pi_0\ ,
\end{array}\right.\ \
\Gamma_m^* = \left\{\begin{array}{ll}
1 & m=\sigma\\
i\tau_-\gamma_5 & m=\pi_+ \\
i\tau_+\gamma_5 & m=\pi_- \\
i\tau_3\gamma_5 & m=\pi_0 \\
\end{array}\right.
\end{equation}
Since the Lorentz invariance is explicitly broken at finite
temperature, the polarization functions depend on $k_0^2$ and
${\bf k}^2$ separately.

With the NJL meson propagator (\ref{meson1}), the quark potential
through a meson exchange is expressed as
\begin{eqnarray}
\label{potential1}
V_m(r) &=& \int{d^3{\bf k}\over
(2\pi)^3}e^{i{\bf k}\cdot{\bf r}}U_m(0,k^2)\nonumber\\
 &=&-{G\over 2\pi^2 r}{\text{Im}} \int_{-\infty}^{\infty} dk
{ke^{ikr}\over 1-2G\Pi_m\left(0,k^2\right)}.
\end{eqnarray}
By constructing a contour in the complex meson momentum plane, the
above integration can be done via calculating the meson pole in
the complex plane. To this end, we need to know the analytic
structure of the meson polarization function $\Pi_m(k_0^2,k^2)$.

In the Pauli-Villars regularization scheme we again divide the
polarization function into a vacuum and a matter part,
\begin{equation}
\label{meson3}
\Pi_m(k_0^2,k^2) =
\Pi_m^{vac}(k_0^2,k^2)+\Pi_m^{mat}(k_0^2,k^2).
\end{equation}
From the discontinuity of the matter part on the real axis and the
discontinuity of the vacuum part on the imaginary
axis~\cite{florkowski1},
\begin{widetext}
\begin{eqnarray}
\label{meson4}
&& \text{Re}\Pi_m^{vac}(0,(k\pm
i\epsilon)^2)=I_1^{vac}+{3(k^2+\epsilon_m^2)\over
2\pi^2}\sum_{j=0}^N A_j\left[{1\over k_j}\sqrt{1+k_j^2}\
\ln\left(\sqrt{1+k_j^2}+k_j\right)+\ln\Lambda_j\right],\nonumber\\
&& \text{Re}\Pi_m^{mat}(0,(k\pm
i\epsilon)^2)=I_1^{mat}-{3(k^2+\epsilon_m^2)\over
2\pi^2k}\sum_{j=0}^N\sum_{a=\pm} A_j \int_0^\infty dp {p\over
E_j(p)}\ \ln\bigg|{k-2p\over
k+2p}\bigg|f\left(E_j^a(p)\right),\nonumber\\
&& \text{Im} \Pi_m^{vac}(0,(k\pm i\epsilon)^2)=0,\nonumber\\
&& \text{Im} \Pi_m^{mat}(0,(k\pm
i\epsilon)^2)=\mp{3T(k^2+\epsilon_m^2)\over
4k\pi}\sum_{j=0}^N\sum_{a=\pm} A_j
\ln\left(1+e^{-E_j^a(k/2)}\right),\nonumber\\
&&
\text{Re}\Pi_m^{vac}(0,(ik\pm\epsilon)^2)=\left\{\begin{array}{ll}
I_1^{vac}+{3(\epsilon_m^2-k^2)\over 2\pi^2}\sum_{j=0}^N
A_j\left[{1\over k_j}\sqrt{1-k_j^2}\
\arcsin k_j+\ln\Lambda_j\right]\ \ \ \ & \textrm{for $k_j<1$}\\
I_1^{vac}+{3(\epsilon_m^2-k^2)\over 2\pi^2}\sum_{j=0}^N
A_j\left[{1\over k_j}\sqrt{k_j^2-1}\
\ln\left(k_j+\sqrt{k_j^2-1}\right)+\ln\Lambda_j\right]\ \ \ \ &
\textrm{for $k_j>1$}\end{array}\right.\nonumber\\
&& \text{Re}\Pi_m^{mat}(0,(ik\pm\epsilon)^2)
=I_1^{mat}-{3(\epsilon_m^2-k^2)\over
2\pi^2k}\sum_{j=0}^N\sum_{a=\pm} A_j \int_0^\infty dp {p\over
E_j(p)}\left[\pi-2\arctan \left({k\over
2p}\right)\right]f\left(E_j^a(p)\right),\nonumber\\
&&
\text{Im}\Pi_m^{vac}(0,(ik\pm\epsilon)^2)=\left\{\begin{array}{ll}
0\ \ \ \ & \textrm{for $k_j<1$}\\
\pm {3(\epsilon_m^2-k^2)\over 4\pi}\sum_{j=0}^N A_j {1\over
k_j}\sqrt{k^2_j-1}\ \ \ \ & \textrm{for
$k_j>1$}\end{array}\right.\nonumber\\
&& \text{Im} \Pi_m^{mat}(0,(ik\pm\epsilon)^2)=0
\end{eqnarray}
\end{widetext}
with the definition of $k_j=k/(2\Lambda_j)$ and $\epsilon_m^2=0$
for $m=\pi_+,\pi_-,\pi_0$ and $\epsilon_m^2=4M^2$ for $m=\sigma$,
there is a vacuum cut on the imaginary axis starting at $2m$ and a
matter cut on the real axis in the complex meson momentum plane.
To replace the direct momentum integration (\ref{potential1}) by
the contour integration in the complex momentum plane, we need to
know the pole position of the meson propagator. Suppose the pole
is at $k=iM_m + \Gamma_m$, the mass $M_m$ and width $\Gamma_m$ are
determined by the complex pole equation
\begin{equation}
\label{pole1}
1-2G\Pi_m\left(0,(iM_m+\Gamma_m)^2\right)=0.
\end{equation}

Since the Lorentz invariance is broken at finite temperature, the
mass $M_m$ and width $\Gamma_m$ related to the quark potential are
in general different from the dynamic meson mass $\overline M_m$
and the corresponding width $\overline\Gamma_m$ controlled by the
pole equation
\begin{equation}
\label{pole2}
1-2G\Pi_m\left((\overline
M_m-i\overline\Gamma_m)^2,0\right)=0.
\end{equation}
Only in the vacuum with $T=\mu=0$, the two poles coincide.

Suppose the width is much less than the mass, $\Gamma_m<<M_m$, the
mass is decoupled from $\Gamma_m$,
\begin{equation}
\label{mass}
1-2G\ \text{Re}\Pi_m(0,(iM_m+\epsilon)^2)=0,
\end{equation}
and $\Gamma_m$ is determined by
\begin{equation}
\label{width}
\Gamma_m=- {\epsilon_m^2-M_m^2\over
2M_m}{\text{Im}\Pi_m(0,(iM_m+\epsilon)^2)\over\text{Re}\Pi_m(0,(iM_m+\epsilon)^2)-I_1}.
\end{equation}
Due to the mass constraint $\overline M_\pi<2M$ at $T<T_c$, pions
are bound states and can not decay into two quarks, and therefore
they have no width in the chiral breaking phase. In chiral limit
with zero current quark mass, from the mass relations
$M_m=\epsilon_m$ at $T<T_c$, not only pions but also sigma have no
widths in the chiral breaking phase.

Taking into account the contributions from the vacuum cut, matter
cut and the pole to the total quark potential, we have
\begin{equation}
\label{potential2}
V_m(r)=V_m^{vac}(r)+V_m^{mat}(r)+V_m^{pol}(r)
\end{equation}
with the vacuum part
\begin{eqnarray}
\label{vc}
V_m^{vac}(r) &=& {G\over 2\pi^2 r}\int_{2M}^\infty dk
ke^{-kr}\text{Im}\bigg[{1\over
1-2G\Pi_m(0,(ik+\epsilon)^2)}\nonumber\\
&&-{1\over 1-2G\Pi_m(0,(ik-\epsilon)^2)}\bigg]
\end{eqnarray}
which decays exponentially, the matter part
\begin{eqnarray}
\label{vm}
V_m^{mat}(r) &=& -{G\over 2\pi^2
r}\int_{-\infty}^\infty dk k\ {\text {Im}}\ e^{ikr}\bigg[{1\over
1-2G\Pi_m(0,k^2)}\nonumber\\
&&-{1\over 1-2G\Pi_m(0,(k+i\epsilon)^2)}\bigg]
\end{eqnarray}
which in general oscillates, and the pole part
\begin{equation}
\label{vp}
V_m^{pol}(r) = {1\over 2\pi} {e^{-M_m r}\over
r}{a_m\cos(\Gamma_m r)+b_m\sin(\Gamma_m r)\over a_m^2+b_m^2}
\end{equation}
with the coefficients $a_m$ and $b_m$ defined by
\begin{equation}
\label{ambm}
{\partial\Pi_m\left(0,k^2\right)\over\partial
k^2}\bigg|_{k=\Gamma_m+iM_m}=a_m+ib_m,
\end{equation}
where the mass and width control, respectively, the amplitude and
the oscillation frequency of the pole contribution. In the chiral
breaking phase, the pole is exactly on the imaginary axis, we
recover the traditional Yukawa form
\begin{equation}
\label{yukawa}
V_m^{pol}(r) ={1\over 2\pi} {e^{-M_m r}\over a_m
r}.
\end{equation}

\section {results}
\label{s3}

\begin{figure}[!htb]
\begin{center}
\includegraphics[width=8cm]{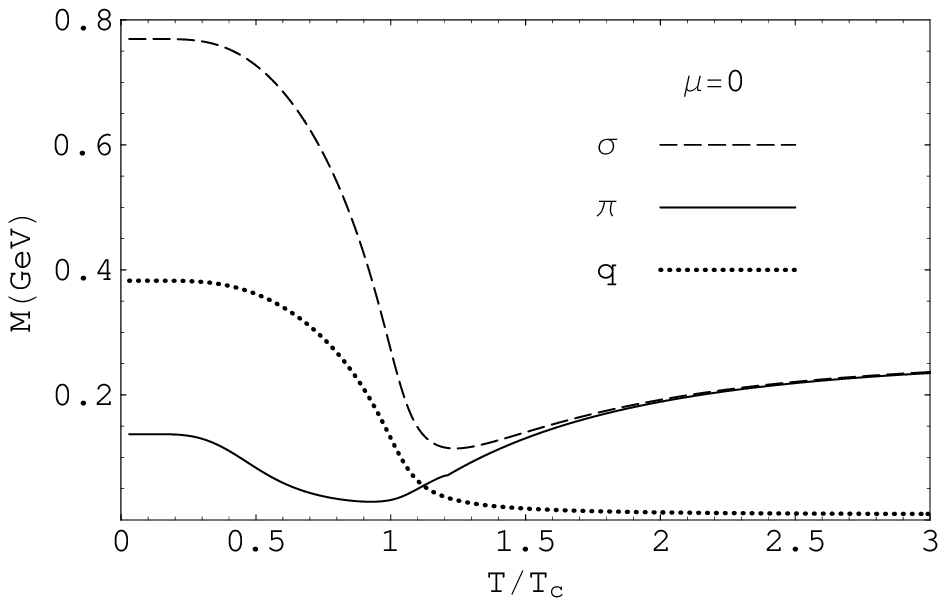}
\includegraphics[width=8cm]{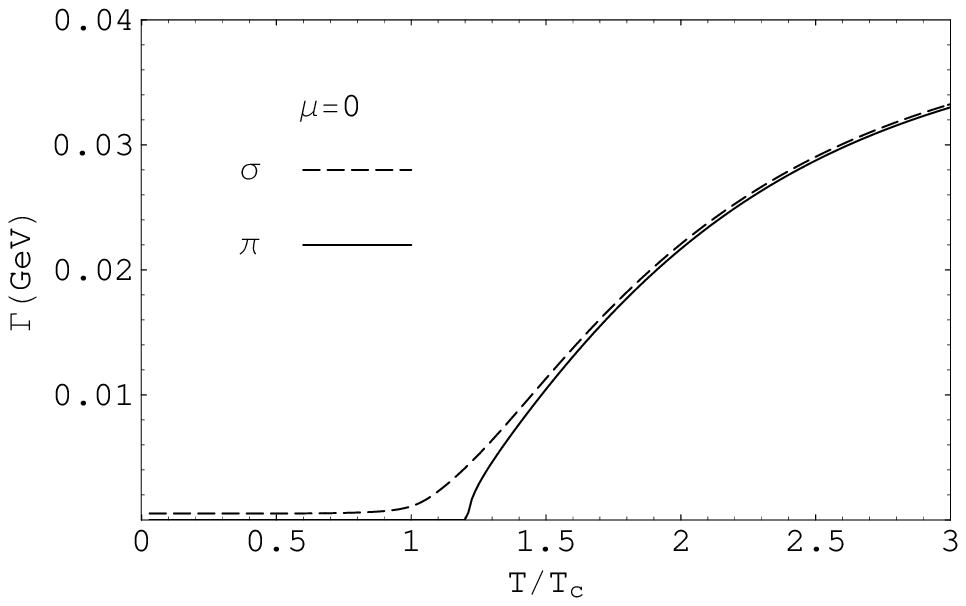}
\caption{The quark mass, meson masses (upper panel) and meson
widths (lower panel) as functions of temperature at zero chemical
potential. The temperature is scaled by the critical value $T_c$.}
\label{fig2}
\end{center}
\end{figure}

We first show the effect of chiral symmetry restoration on the
quark potential in hot and dense medium. With the parameter values
in the model and in the Pauli-Villars regularization scheme, shown
in the last section, the pole mass $M_m$ and width $\Gamma_m$ as
functions of temperature at zero chemical potential are presented
in Fig.\ref{fig2} for $m=\pi,\ \sigma$. In chiral limit, the
critical temperature $T_c$ for chiral phase transition is about
$300$ MeV. With different parameter values or choosing other
regularization schemes, the critical temperature can be reduced.
For instance, in the three-momentum non-covariant cutoff scheme,
the critical value is $T_c=170$ MeV~\cite{zhuang}. In the real
world with nonzero current quark mass, pions and sigma behave
differently in the chiral breaking phase. Pions are bound states
with zero width, but sigma mass is slightly larger than two times
the quark mass. Therefore, there is a small width with sigma at
low temperature. When chiral symmetry is restored at high
temperature $T>T_c$, the difference between pions and sigma
disappears, both the masses and widths coincide. The width is
always much smaller than the corresponding mass. Even at high
temperature $T/T_c=3$ the ratio $\Gamma_m/M_m$ is only about
$10\%$. This is in consistent with the assumption of $\Gamma_m <<
M_m$ for the decoupling of $M_m$ from $\Gamma_m$, (\ref{mass}) and
(\ref{width}). The meson masses get saturated when the temperature
is high enough, this will lead to the saturation of the quark
potential, see the following discussion. Note that, the finite
width plays a crucial rule in the chiral restoration phase. At
$T>T_c$, quarks become almost massless, and pions and sigma
satisfy the decay condition $M_m>2M$. If the meson width is
neglected, the pole is located on the imaginary axis and goes into
the vacuum cut, and the pole contribution will be fully
suppressed. However, for a nonzero width, even if it is very small
compared with the mass, the pole position deviates from the
imaginary axis, it is outside the vacuum cut and its contribution
is not removed from the quark potential.
\begin{figure}[!htb]
\begin{center}
\includegraphics[width=8cm]{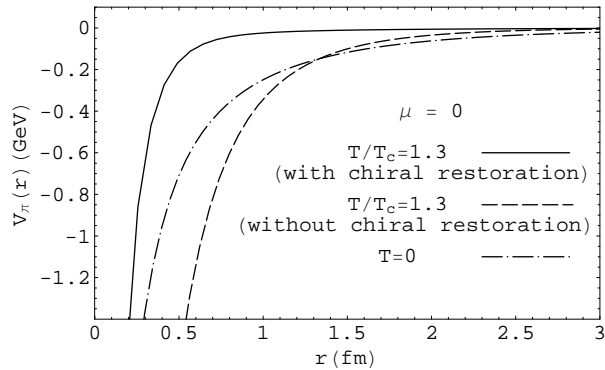}
\caption{The pion-mediated quark potentials with and without
considering chiral symmetry restoration. The results are compared
with the potential in the vacuum. } \label{fig3}
\end{center}
\end{figure}

To directly see the effect of chiral symmetry restoration on the
quark potential, we show in Fig.\ref{fig3} the pion-mediated
potentials at $T/T_c=1.3$ and $\mu=0$ in the cases of with and
without chiral symmetry restoration, and compare them with the
potential in the vacuum ($T=\mu=0$). For the calculation without
chiral phase transition, we keep the quark mass and pion mass as
their values in the vacuum and remain only the explicit
temperature dependence in the distribution function $f$. In this
case, the quark potential becomes stronger in comparison with the
vacuum potential, because of the increasing quark density in hot
medium, and is significantly different from the calculation with
chiral symmetry restoration where the potential is much weaker
than the vacuum potential, due to the change in the chiral
dynamics.

\begin{figure}[!htb]
\begin{center}
\includegraphics[width=8cm]{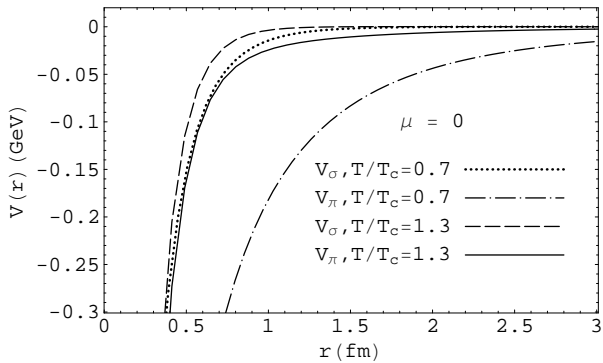}
\caption{The pion- and sigma-mediated quark potentials in the
chiral breaking and chiral restoration phases.} \label{fig4}
\end{center}
\end{figure}

Since sigma is much heavier than pion in the phase with chiral
symmetry breaking, and they become degenerate when the symmetry is
restored, sigma exchange is remarkably reflected in the quark
potential only in the short range for $T<T_c$ and becomes
equivalently important as pion exchange for $T>T_c$. This is
explicitly shown in Fig.\ref{fig4}.
\begin{figure}[!htb]
\begin{center}
\includegraphics[width=8cm]{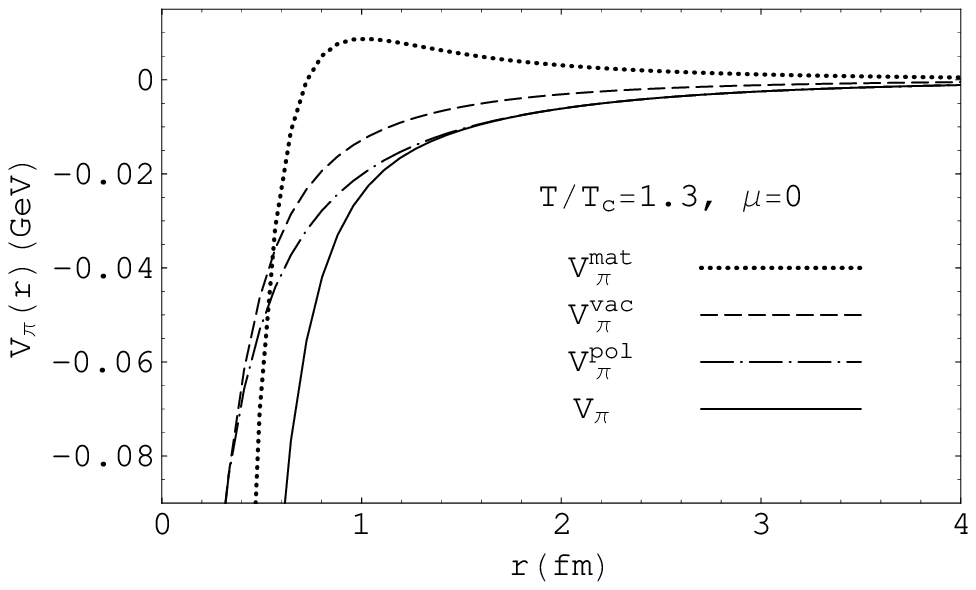}
\includegraphics[width=8cm]{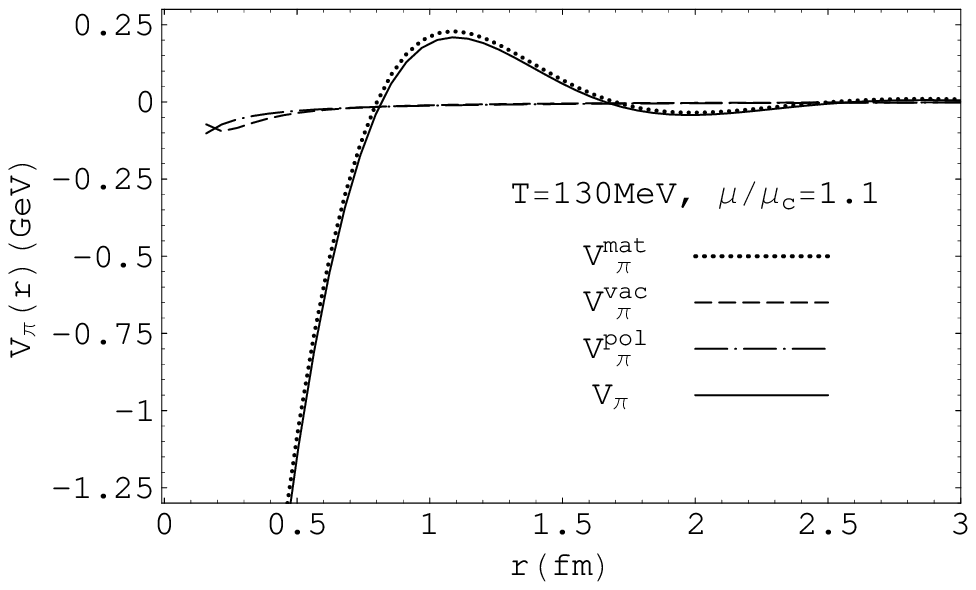}
\caption{The pion-mediated quark potential and its three
components arising from the vacuum cut, matter cut, and pole in
the chiral restoration phase at high temperature ($\mu=0$ and
$T/T_c=1.2$, upper panel) and high chemical potential ($T=130$ MeV
and $\mu/\mu_c=1.1$, lower panel).} \label{fig5}
\end{center}
\end{figure}

We now discuss the two possible oscillations in the quark
potential, the Friedel oscillation and the Yukawa oscillation. The
former is due to the sharp Fermi surface and the latter arises
from complex poles of meson exchanges. From the matter and pole
contributions (\ref{vm}) and (\ref{vp}), there are always
oscillations in the two parts. To know whether they can be clearly
reflected in the total potential, we separately show the three
contributions from the vacuum cut, matter cut and pole in
Fig.\ref{fig5}. While there is a remarkable oscillation around
$r=1$ fm in the matter part at $\mu=0$ and $T/T_c=1.3$, see the
upper panel, it can not be seen clearly in the total potential,
since at high temperature and low density the matter contribution
is not the dominant one in the total potential. In this case, only
Yukawa oscillation can be seen, if the width is large enough. At
$T/T_c=1.3$ the pion width is too small and the induced weak
oscillation is difficult to be demonstrated in the total
potential. On the other hand, at low temperature and high density,
the quark potential is absolutely controlled by the matter cut,
see the lower panel at $\mu/\mu_c=1.1$ with $\mu_c=400$ MeV being
the critical chemical potential for $T=130$ MeV, and only the
Fermi-surface induced Friedel oscillation can be seen in the total
potential. When the temperature and chemical potential are not low
enough, especially in the chiral restoration phase, the
contribution from the vacuum cut is always smaller than the matter
part and pole part.

\begin{figure}[!htb]
\begin{center}
\includegraphics[width=8cm]{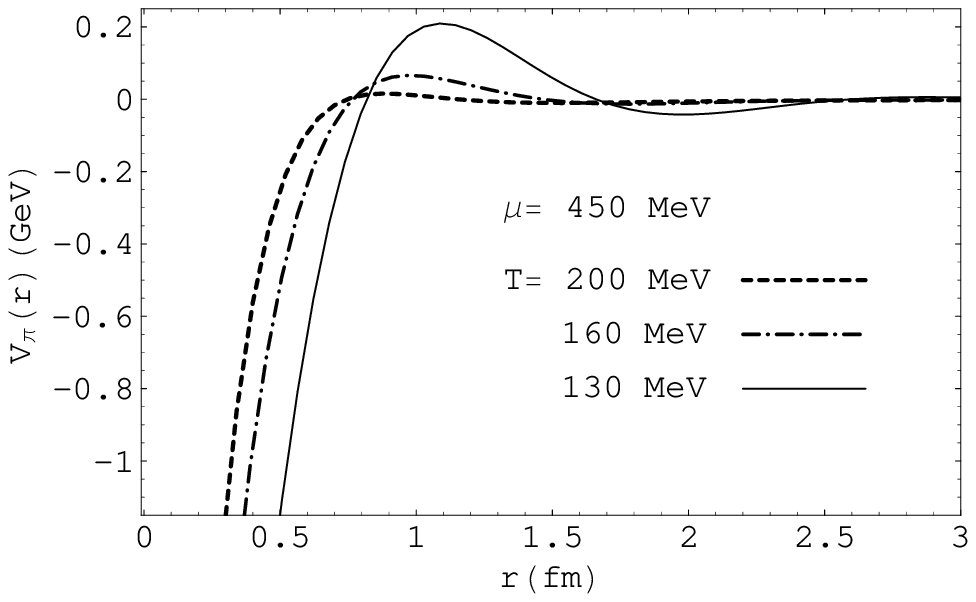}
\includegraphics[width=8cm]{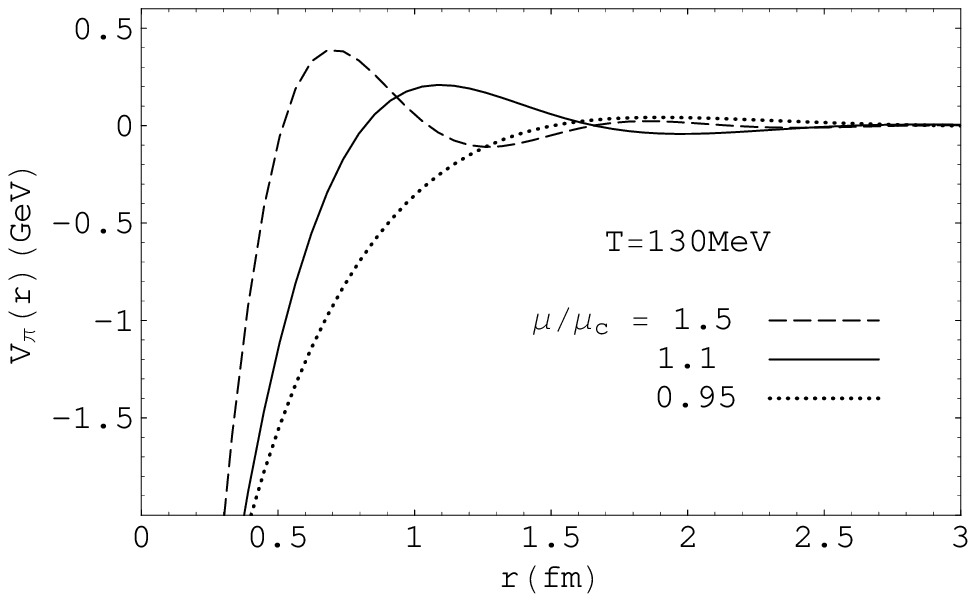}
\caption{The temperature (upper panel) and chemical potential
(lower panel) dependence of the Friedel oscillation in the
pion-mediated quark potential. }
\label{fig6}
\end{center}
\end{figure}

The Friedel oscillation is related to the singularity of the
matter part of the meson polarization function $\Pi_m$ at $k=2p_f$
on the real axis at zero temperature, where $p_f$ is the quark
Fermi momentum, see (\ref{meson4}). At finite temperature, there
is no more strict Fermi surface and singularity in the
polarization function, the oscillation will become weak at low
temperature and finally disappear at high enough temperature. In
the upper panel of Fig.\ref{fig6} we plot the quark potential at
fixed chemical potential for three values of temperature. The
Friedel oscillation is gradually washed away by increasing
temperature. The chemical potential dependence of the Friedel
oscillation at fixed temperature is shown in the lower panel. In
the chiral breaking phase with low chemical potential $\mu<M$,
there is no Fermi momentum and in turn no Friedel oscillation. The
oscillation happens only in the chiral restoration phase, and the
amplitude and frequency increase with increasing chemical
potential.
\begin{figure}[!htb]
\begin{center}
\includegraphics[width=8cm]{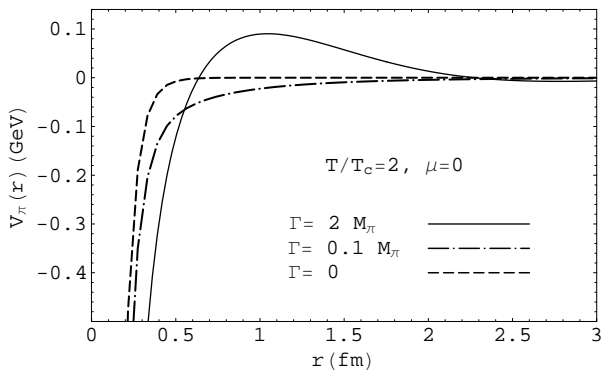}
\caption{The width dependence of the Yukawa oscillation in the
pion-mediated quark potential in the chiral restoration phase. }
\label{fig7}
\end{center}
\end{figure}

The strength of the Yukawa oscillation depends strongly on the
pole width. In Fig.\ref{fig7} we show the pion-mediated quark
potential in the chiral restoration phase with $\mu=0$ and
$T/T_c=2$ for three different values of width. As we emphasized
above, the potential at high temperature and low density is
dominated by the pole contribution. From the analytic structure of
the meson polarization function (\ref{meson4}), its matter part is
continuous on the imaginary axis, and only the vacuum part
contributes to the width (\ref{width}). From our numerical
calculation, see Fig.\ref{fig2}, the width is only about $10\%$ of
the mass at $T/T_c=2$, and therefore the Yukawa oscillation can
not be clearly shown in Fig.\ref{fig7}. However, this small width
can not be neglected, otherwise the pole will be located on the
imaginary axis and go into the vacuum cut, and has no contribution
to the potential. To see a clear Yukawa oscillation, we take
artificially the width as a free parameter and calculate the
potential with $\Gamma_\pi/M_\pi=2$. In this case, there is really
a strong oscillation around $r=1$ fm, as we expected.

\begin{figure}[!htb]
\begin{center}
\includegraphics[width=8cm]{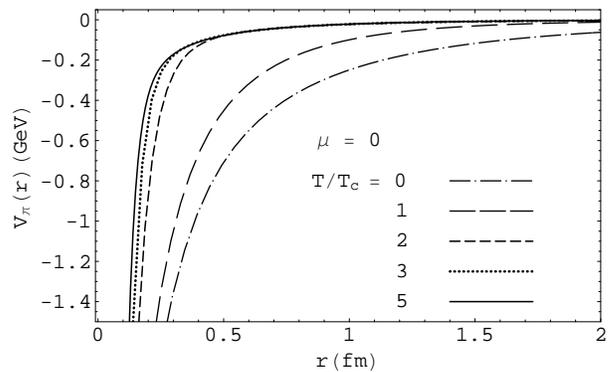}
\caption{The temperature dependence of the pion-mediated quark
potential at zero chemical potential. The potential gets saturated
at $T/T_c=3$. } \label{fig8}
\end{center}
\end{figure}

From the analysis of the experimental data obtained in
relativistic heavy ion collisions, especially the anisotropic
flow, and the lattice and model calculation on the resonant states
of quarks and gluons above the critical temperature, it is widely
accepted that the parton system above and close to the critical
temperature is a strongly coupled quark-gluon plasma. The
temperature region is about $1 \le T/T_c <
3\sim4$~\cite{shuryak3}. Above this region, quarks and gluons are
weakly coupled and the system can approximately be considered as
an ideal gas. In Fig.\ref{fig8} we show the temperature dependence
of the pion-mediated quark potential at zero chemical potential.
With increasing temperature, the potential becomes weaker and
weaker due to the chiral symmetry restoration, and finally gets
saturated for $T/T_c\gtrsim 3$. Since the used Pauli-Villars
regularization scheme guarantees the ideal gas limit at
$T\rightarrow\infty$~\cite{florkowski1}, we can conclude from this
saturation that, the quark-meson plasma is strongly coupled in the
temperature region $1\le T/T_c\lesssim 3$. Above $T/T_c\simeq 3$,
the system behaves like an ideal gas. While there are no gluons
and confinement in the NJL model, the quark-meson plasma has the
similar strong coupling region as the quark-gluon plasma.

\begin{figure}[!htb]
\begin{center}
\includegraphics[width=8cm]{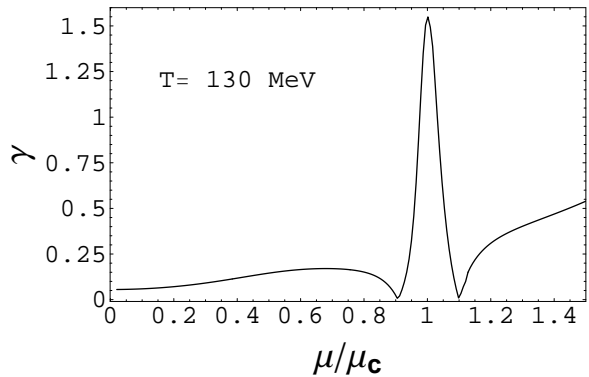}
\caption{The ratio of quark potential to quark kinetic energy as a
function of chemical potential at $T=130$ MeV. The peak is located
at the critical point of chiral symmetry restoration.}
\label{fig9}
\end{center}
\end{figure}

To describe the degree of particle coupling in matter, one can
introduce the ratio of particle potential to particle kinetic
energy~\cite{shuryak4},
\begin{equation}
\label{ratio}
\gamma=V_0/E_{kin}.
\end{equation}
A large ratio means a strongly coupled matter and a small ratio
indicates a weakly coupled matter. For the considered quark-meson
plasma, we take the potential $V_0$ as the meson-mediated quark
potential at the average distance between two quarks,
\begin{equation}
\label{av}
V_0 = |V(2r_0)|,
\end{equation}
where $V$ is the summation of $V_m$ over all possible meson
exchanges, the averaged quark radius $r_0$ is determined by the
quark and anti-quark number densities $n_q$ and $n_{\bar q}$,
\begin{eqnarray}
\label{r0}
&& \left(n_q+n_{\bar q}\right){4\over 3}\pi r_0^3=1,\nonumber\\
&& n_{q,\bar q}(T,\mu)=\int {d^3{\bf p}\over
(2\pi)^3}f\left(E(p)\mp\mu\right),
\end{eqnarray}
with $E(p)=\sqrt {{\bf p}^2+M^2}$, and the quark kinetic energy is
\begin{eqnarray}
\label{kin} && E_{kin}= {1\over n_q+n_{\bar q}}\int {d^3{\bf
p}\over (2\pi)^3}\left(E(p)-M\right)\nonumber\\
&&\ \ \ \ \ \ \ \ \ \ \times\left(f(E(p)-\mu)+f(E(p)+\mu)\right).
\end{eqnarray}
The ratio is shown in Fig.\ref{fig9} as a function of chemical
potential at fixed temperature $T=130$ MeV. Different from the
kinetic energy $E_{kin}$ which monotonously goes up with
increasing temperature or chemical potential, the potential $V_0$
depends on both the shape of the potential and the quark density.
The former is controlled by chiral dynamics and the latter is
determined by thermodynamics. With increasing temperature or
chemical potential, the quark potential becomes weak, while the
density increases and in turn the radius $r_0$ gets small. From
our numerical calculation, the ratio is rather small at low and
high chemical potentials, but peaks strongly at the critical
point. This tells us that the strongest coupling in the
quark-meson plasma is located at the chiral phase transition. The
smoothly increasing ratio after the phase transition is due to the
strong Friedel oscillation at high density. From the AdS/CFT
estimation~\cite{adscft} and perturbative QCD
calculation~\cite{pqcd}, it is well-known that the ratio of shear
viscosity to entropy $\eta/s$ has the minimum value at the
critical point of any phase transition, which is widely considered
as a general property of strongly coupled matter. The maximum of
the ratio $V_0/E_{kin}$ at the chiral phase transition agrees well
with the minimum of the ratio $\eta/s$, both describe the
strongest coupling at the phase transition.

\section {Conclusions}
\label{s4}

In the frame of the two flavor SU(2) NJL model, we calculated the
light quark potential through pion and sigma exchanges in RPA
approximation. The chiral symmetry restoration at high temperature
and density plays an essential rule in the quark potential. While
the matter part is always associated with an oscillation, it
dominates the potential only for dense matter, and therefore the
oscillation can be reflected clearly in the total quark potential
only at low temperate and high density. The other possible
oscillation in the chiral restoration phase is the Yukawa
oscillation induced by the complex poles of mesons. From our
self-consistent calculation for meson mass and width, the Yukawa
oscillation is weak even at extremely high temperature. The
potential gets saturated when temperature is larger than about
three times the critical temperature. This means that the strongly
coupled quark-meson plasma is in the region of $1\le T/T_c\lesssim
3$, which agrees well with the widely accepted result for the
strongly coupled quark-gluon plasma. Corresponding to the
well-known minimum of the ratio of shear viscosity to entropy, we
found that the ratio of potential to kinetic energy peaks strongly
at the critical point of chiral phase transition.

While we found the strong coupling region for the quark-meson
plasma through considering the saturation of quark potential, the
saturation temperature $T/T_c=3$ looks beyond the effective range
of the NJL model. The reliable conclusion is probably that the
quark-meson plasma above and close to the chiral critical point is
strongly coupled. In our calculation we did not include the vector
meson exchange which is expected to contribute a repulsive part to
the quark potential and then reduce the depth of the potential
well in the short range. To have a complete calculation of the
quark potential in the NJL model, one should consider the vector
mesons. Another point is the deconfinement effect, it will be
helpful to investigate the quark potential using the improved NJL
model with Polyakov loop dynamics~\cite{pnjl1, pnjl2}. Since the
original NJL model with pions and sigma describes well the chiral
dynamics in the vacuum and at finite temperature and density, we
hope our result on the quark potential in the quark-meson plasma
is helpful to understand the effect of chiral symmetry restoration
on the strongly coupled quark-gluon plasma.
\\ \\
{\it Acknowledgements:} We have benefited from discussions with W.
Florkowski. P.Z. thanks the Yukawa Institute for Theoretical
Physics at Kyoto University and the organizers of NFQCD2008 for
their hospitality and for providing the atmosphere of stimulating
discussions. The work is supported by the NSFC Grants 10575058 and
10735040 and the 973-project 2007CB815000.



\end{document}